\documentclass{sig-alternate}
\usepackage[numbers,sort]{natbib}
\usepackage{url}
\usepackage{verbatim}
\usepackage{amsfonts}
\usepackage{bm}
\usepackage{amssymb}
\usepackage{amsmath}
\usepackage{xspace}
\usepackage[center]{subfigure} 
\usepackage{caption}
\usepackage{mathtools}
\usepackage{mathptmx} 
\usepackage{microtype} 

\usepackage{setspace}

\usepackage{color,colortbl}
\definecolor{Gray}{gray}{0.9}
\definecolor{LightCyan}{rgb}{0.88,1,1}
\definecolor{Pink}{rgb}{1,0.75,0.70}
\definecolor{Gray}{rgb}{0.84,0.84,0.84} 

\newcommand{\para}[1]{\smallskip\noindent\textbf{#1}} 

\begin{document}






\permission{Copyright is held by the author/owner(s).}
\conferenceinfo{WWW'16 Companion,}{April 11--15, 2016, Montr\'{e}al, Qu\'{e}bec, Canada.}
\copyrightetc{ACM \the\acmcopyr}
\crdata{978-1-4503-4144-8/16/04. \\
http://dx.doi.org/10.1145/2872518.2889385 }


\clubpenalty=10000
\widowpenalty = 10000

%


\title{Inferring Gender from Names on the Web: \\A Comparative Evaluation of Gender Detection Methods}


%
%
%
%

\numberofauthors{1}
\author{
   \alignauthor {Fariba Karimi\textsuperscript{*}, Claudia Wagner\textsuperscript{*}, Florian Lemmerich\textsuperscript{*}, \\Mohsen Jadidi\textsuperscript{*}, and Markus Strohmaier\textsuperscript{* \dag}}\\
     \affaddr{\textsuperscript{*} GESIS - Leibniz Institute for the Social Sciences}\\
     \affaddr{\textsuperscript{\dag} University of Koblenz-Landau}\\
     \email{\{firstname.lastname\}@gesis.org}
}


\maketitle
\begin{abstract}

Computational social scientists often harness the Web as a ``societal observatory'' where data about human social behavior is collected. This data enables novel investigations of psychological, anthropological and sociological research questions.
However, in the absence of demographic information, such as gender, many relevant research questions cannot be addressed. To tackle this problem, researchers often rely on automated methods to infer gender from name information provided on the web. However, little is known about the accuracy of existing gender-detection methods and how biased they are against certain sub-populations. In this paper, we address this question by systematically comparing several gender detection methods on a random sample of scientists for whom we know their full name, their gender and the country of their workplace.
We further suggest a novel method that employs web-based image retrieval and gender recognition in facial images in order to augment name-based approaches. Our findings show that the performance of name-based gender detection approaches can be biased towards countries of origin and such biases can be reduced by combining name-based an image-based gender detection methods.


\end{abstract}



\section{Introduction}
\label{sec:introduction}
The Web enables studies of human social behavior on a very large scale. For many research questions, demographic information about individuals (such as age, gender or ethnic background) is highly beneficial but often particularly difficult to obtain.


This has led previous research to employ different methods for inferring the gender of individuals from names. For example, in \cite{green2009gender} the authors determine the gender of individuals using the name repository from the US Social Security Administration and study the relationship between gender and job performance among brokerage firm. Mislove et al.~\cite{mislove2011understanding} used the same name repository to infer the gender of Twitter users by mapping their self-reported names to the name database. In another study the authors aim to study gender disparities in science and infer the gender of scientists based on a similar approach \cite{West2013}. 
Unfortunately, most previous work does not provide information on how accurate different gender detection methods are and/or how biased they are against certain sub-populations. Although crowd sourcing methods can be seen as an alternative for automated gender detection methods, they do not scale well and are expensive. In the absence of a full name, more sophisticated methods such as supervised machine learning models are used to harness the users content for detecting the gender (see e.g. \cite{Rao2010}). Yet, separate models are needed for gender detection methods in each language community \cite{ciot2013gender}. 


In this paper we evaluate and compare frequently used name-based gender detection methods. We report overall accuracy and also bias, i.e., deviating accuracy for different demographic sub-populations (e.g. men, women and people living in different countries). Moreover, we propose novel methods that increase the accuracy of gender detection across heterogeneous sub-populations by augmenting traditional methods with face recognition techniques.


\section{Data and method}
\label{sec:data}
For our evaluation, we utilized ground truth data from a previous study on global gender disparities in science~\cite{Cassidy2013}.
It consists of a manually labeled random sample of academics, their full names, institutions, countries, and their gender. 
The ground truth was created by inspecting CVs, pictures and institutional websites. 
After removing ambiguous and repetitive names, the final name list consist of 693 male names and 723 female names. 

We then evaluate different name-based gender detection methods using the full names of our manually labeled scientists as input. Finally, we propose a new mixed method that combines name-based and image-based gender detection. 


\subsection{Gender detection methods}
In the following, we review some prominent unsupervised approaches that only require a name or picture as an input. These approaches do not require training and are widely used in scientific research as mentioned in the introduction.

\smallskip





\para{ Security Administration's baby names data.}
The US Social Security Administration (SSA) covers registered baby names in the United States since 1880. Many gender detection tools such as the ``gender'' package in \textit{R}\protect\footnote{\url{https://cran.r-project.org/web/packages/gender/}} or the OpenGenderTracking\protect\footnote{\url{http://opengendertracking.github.io/}} rely on this database.


\para{IPUMS Census data.}
Integrated Public Use Microdata Series (IPUMS) census data  consists of samples of the American population drawn from fifteen federal censuses from the American Community Surveys between 1850 to 2000. This database is also used in the ``gender'' package in \textit{R} and other web-based name extraction packages\footnote{\url{https://usa.ipums.org/usa-action/}}. 

\para{Sexmachine.} The list of 40,000 names is primarily collected by J\"org Michael. Because of its availability, several libraries in various languages (see for example C's (\textit{gender.c}) or Python's \textit{Sexmachine} library\footnote{\url{https://pypi.python.org/pypi/SexMachine/}}) use this database. Given a name, Sexmachine makes a guess whether the name is male, mostly male, female, mostly female or unclear. The advantage of this name list is that it provides detailed information about how popular a first name is in a country and how strongly it is associated with a given gender. Therefore, it enables the disambiguation of names based on the country of origin. The list also provides information for a variety of countries including China and India.

\para{Genderize.} Apart from publicly available name data bases there are numerous commercial applications that incorporate various databases from online resources to assess gender. The problem with commercial applications is the difficulty to determine how the data is gathered and processed. Among commercial detection methods are \textit{Facebook graph API}, \textit{Gender API}, \textit{Namsor} which is based on \textit{Gender API} and \textit{Genderize}. In this work, we analyzed the latter method. 
\textit{Genderize} utilizes big datasets of information, from user profiles across major social networks and exposes this data through its API. The response includes a confidence value\footnote{\url{https://genderize.io/}}.

\para{Face++.} In addition to name-based gender detection methods, face recognition algorithms have become a popular tool for inferring the gender, e.g., for social media users. Among those, image-based application \textit{Face++} seems to provide high performance \cite{zhou2015naive}. This approach requires access to a picture of the person.

In order to derive the gender for a specific scientist, we propose to initially collect the first five \textit{Google thumbnails} using the full name as search query term and then apply image-recognition on the search results.
This approach does not necessarily require that the collected pictures depict the scientist we originally searched for, but the idea is that we collect a sample of pictures that depict people who are named like the person we searched for. The advantage of using the full name as input is that for first names that are ambiguous or unisex, the combination of first and last name is often a better indicator of the gender associated with the certain culture.

\para{A Novel Mixed Approach.} 
In addition, we propose mixed methods that combine name-based detection methods with an image-based face recognition approach. We test two variations of this method. In method \textit{Mixed1}, the best name-based approach, namely \textit{Genderize}, is used first. For the remaining unidentified names, the image-based method \textit{Face++}, is used. In method \textit{Mixed2}, \textit{Genderize} and \textit{Face++} have equal weight. For the weighting, we do not use a binary decision for each method, but also take the reported confidence as a numeric value into account. In doing so, this method can handle ambiguous names more efficiently. Note that method \textit{Mixed1} does not require retrieving pictures for the whole population and is therefore more efficient than method \textit{Mixed2}.

\begin{table}[]
\centering
\caption{Per-class and overall precision and recall of various gender detection methods. The mixed approach outperforms all other methods by at least 9\%.}
\label{tab:results}
\resizebox{\columnwidth}{!}{%
\begin{tabular}{|l|l|l|l|l|l|l|l|}
\hline
                 & SSA  & IPUMS & Sexmachine & Genderize &Face++  &Mixed1 & Mixed2 \\ 
\hline
\rowcolor{Pink}
female precision & 0.96 & 0.96  & 0.97       & 0.95      & 0.92     &0.91     & 0.93      \\ 
\rowcolor{Pink}
female recall    & 0.79 & 0.69  & 0.77       & 0.86      & 0.81     &0.95    & 0.94      \\ 
\rowcolor{Pink}
female $F_1$     &0.86  &0.80   &0.85        &0.90       &0.86      &0.93    &0.93       \\ \hline

\rowcolor{LightCyan}
male precision   & 0.98 & 0.92  & 0.98       & 0.98      & 0.86     &0.96   & 0.98      \\ 
\rowcolor{LightCyan}
male recall      & 0.70 & 0.68  & 0.72       & 0.77      & 0.85     &0.89   & 0.88      \\ 
\rowcolor{LightCyan}

male $F_1$      &0.82   &0.78   &0.83       &0.86       &0.85       &0.92   &0.93 \\ \hline
accuracy         & 0.75 & 0.68  & 0.74       & 0.82      & 0.83  &\cellcolor{Gray}0.92     & 0.91      \\ \hline
\end{tabular}
}

\end{table}

\begin{table}[h!]
\centering
\caption{Accuracy of various gender detection methods for people from different countries. For most countries mixed approaches perform best.}
\label{tab:accuracy_country}
\resizebox{\columnwidth}{!}{%
\begin{tabular}{|l|c|c|c|c|c|c|c|c|}
\hline
               & \# instances & SSA  & IPUMS & Sexmachine              & Genderize & Face++                 & Mixed1               & Mixed2                  \\ \hline
United States  & 419         & 0.82 & 0.76  & 0.84                    & 0.83      & 0.91                    & \cellcolor{Gray} 0.91 & 0.90                    \\ \hline
China          & 113         & 0.20 & 0.11  & \cellcolor{Gray} 0.67 & 0.28      & 0.65                      & 0.50                    & 0.56                    \\ \hline
United Kingdom & 96          & 0.94 & 0.92  & 0.92                    & 0.94      & 0.81                    & \cellcolor{Gray} 0.98 & 0.94                    \\ \hline
Germany        & 82          & 0.87 & 0.88  & \cellcolor{Gray} 0.96 & 0.94      & 0.87                      & \cellcolor{Gray} 0.96 & 0.93                    \\ \hline
Italy          & 75          & 0.93 & 0.92  & 0.94                    & 0.98      & 0.79                    & 0.99                    & \cellcolor{Gray} 1    \\ \hline
Canada         & 60          & 0.87 & 0.77  & 0.86                    & 0.91      & 0.90                    & \cellcolor{Gray} 0.96 & 0.93                    \\ \hline
France         & 58          & 0.93 & 0.92  & 0.80                    & 0.96      & 0.81                    & 0.97                    & \cellcolor{Gray} 1    \\ \hline
Japan          & 56          & 0.79 & 0.70  &\cellcolor{Gray} 1                    & 0.90      & 0.62                    & 0.91                    & \cellcolor{Gray} 0.94 \\ \hline
Brazil         & 44          & 0.29 & 0.29  & 0.15                    & 0.44      & 0.81                    & 0.90                    & \cellcolor{Gray} 0.93 \\ \hline
Spain          & 39          & 0.96 & 0.92  & 0.92                    &\cellcolor{Gray} 1      & 0.92                    & \cellcolor{Gray} 1    & \cellcolor{Gray} 1    \\ \hline
Australia      & 31          & 0.89 & 0.89  & 0.90                    & 0.86      & 0.86                    & \cellcolor{Gray} 0.94 & 0.93                    \\ \hline
India          & 29          & 0.67 & 0.17  & 0.71                    & 0.78      & 0.83                    & 0.83                    & \cellcolor{Gray} 0.93 \\ \hline
South Korea    & 27          & 0.04 & 0.00  & 0.58                    & 0.11      & \cellcolor{Gray} 0.74   & 0.37                    & 0.66                    \\ \hline
Switzerland    & 25          & 0.78 & 0.70  & 0.56                    & 0.83      & 0.88                    & 0.90                    & \cellcolor{Gray} 0.92 \\ \hline
Turkey         & 21          & 0.43 & 0.14  & 0.79                    & 0.81      & 0.86                    & \cellcolor{Gray} 1    & \cellcolor{Gray} 1    \\ \hline

\end{tabular}

}
\end{table}

\section{Results and Discussion}
\label{sec:results}
The results displayed in Table~\ref{tab:results} show that among individual methods, image-based \textit{Face++} and \textit{Genderize} perform relatively better than others. 
However, the overall best results are achieved by the mixed approaches, which outperform all others by at least 8\% accuracy. 
Although all evaluated methods achieve high overall precision, recall rates vary. 
All gender detection methods show comparable results for both classes (male and female) and therefore no systematic gender-bias can be asserted.

By contrast, Table~\ref{tab:accuracy_country} indicates that the error rates strongly depend on the country of residence of an individual. 
While name-based approaches work quite well for western industrialized countries, their performance deteriorates for emerging nations such as China, South Korea or Brazil. Clearly, popular names of these countries are not covered sufficiently in the databases at this point in time. For these countries, an image-based approach leads to substantially better results (e.g., for South Korea the accuracy of image-based approaches is at least 16\% better than the best name-based method). The \textit{Genderize} method that also harnesses social media performs poorly for China, presumably due to accessibility to the Chinese social networking websites. Our proposed mixed approaches outperform the existing methods for the majority of the countries. 

\textbf{Conclusions:} Our results suggest that the performance of name-based gender detection approaches varies according to the country of origin and that performance for emerging nations is particularly weak. Significant enhancements can be achieved by combining name-based with image-based gender detection methods. In the future, our findings could be combined with machine learning approaches to develop better methods for assessing demographic attributes of users on the Web.









\scriptsize 
\raggedright
\sloppy

\bibliographystyle{abbrv}

\begingroup
    \setlength{\bibsep}{1pt}
    \setstretch{1}
    \bibliography{biblio}

\begin{thebibliography}{1}

\bibitem{ciot2013gender}
M.~Ciot, M.~Sonderegger, and D.~Ruths.
\newblock Gender inference of twitter users in non-english contexts.
\newblock In {\em EMNLP}, pages 1136--1145, 2013.

\bibitem{green2009gender}
C.~Green, N.~Jegadeesh, and Y.~Tang.
\newblock Gender and job performance: Evidence from wall street.
\newblock {\em Financial Analysts Journal}, 65(6):65--78, 2009.

\bibitem{Cassidy2013}
V.~Larivi\`{e}re, C.~Ni, Y.~Gingras, B.~Cronin, and C.~R. Sugimoto.
\newblock {Bibliometrics: Global gender disparities in science}.
\newblock {\em Nature}, 504(7479):211--213, Dec. 2013.

\bibitem{mislove2011understanding}
A.~Mislove, S.~Lehmann, Y.-Y. Ahn, J.-P. Onnela, and J.~N. Rosenquist.
\newblock Understanding the demographics of twitter users.
\newblock {\em ICWSM}, 11:5th, 2011.

\bibitem{Rao2010}
D.~Rao, D.~Yarowsky, A.~Shreevats, and M.~Gupta.
\newblock Classifying latent user attributes in twitter.
\newblock In {\em Proceedings of the 2nd International Workshop on Search and
  Mining User-generated Contents}, SMUC '10, pages 37--44, New York, NY, USA,
  2010. ACM.

\bibitem{West2013}
J.~D. West, J.~Jacquet, M.~M. King, S.~J. Correll, and C.~T. Bergstrom.
\newblock The role of gender in scholarly authorship.
\newblock {\em PLoS ONE}, 8(7):e66212, 07 2013.

\bibitem{zhou2015naive}
E.~Zhou, Z.~Cao, and Q.~Yin.
\newblock {Naive-deep face recognition: Touching the limit of LFW benchmark or
  not?}
\newblock {\em arXiv preprint arXiv:1501.04690}, 2015.

\end{thebibliography}
\endgroup

\end{document}